\documentclass[prl,twocolumn]{revtex4}
\usepackage{bm}
\usepackage{amsmath}
\usepackage{amssymb}
\usepackage{graphicx}

\def\braket#1{\left\langle#1\right\rangle}

\begin{document}

\title{Reflection-Free One-Way Edge Modes in a Gyromagnetic Photonic Crystal}

\author{Zheng~Wang}
\author{Y.~D.~Chong}
\author{John D.~Joannopoulos}
\author{Marin~Solja\v{c}i\'{c}}

\affiliation{Department of Physics, Massachusetts Institute of
  Technology, Cambridge, Massachusetts 02139}

\date{\today}

\begin{abstract}
We point out that electromagnetic one-way edge modes analogous to quantum Hall edge states, originally predicted by Raghu and Haldane in 2D gyroelectric photonic crystals possessing Dirac point-derived bandgaps, can appear in more general settings.  In particular, we show that the TM modes in a gyromagnetic photonic crystal can be formally mapped to electronic wavefunctions in a periodic electromagnetic field, so that the only requirement for the existence of one-way edge modes is that the Chern number for all bands below a gap is non-zero.  In a square-lattice gyromagnetic Yttrium-Iron-Garnet photonic crystal operating at microwave frequencies, which lacks Dirac points, time-reversal breaking is strong enough that the effect should be easily observable.  For realistic material parameters, the edge modes occupy a 10\% band gap.  Numerical simulations of a one-way waveguide incorporating this crystal show 100\% transmission across strong defects, such as perfect conductors several lattice constants wide, larger than the width of the waveguide.
\end{abstract}


\maketitle

Photonic crystals---structures with periodicity comparable to the wavelength of light---possess many unusual and technologically important optical properties \cite{Sajeev,Yablonovitch,JJ}.  The theory of these devices relies on an analogy between Maxwell's equations in a periodic medium and quantum mechanics with a periodic Hamiltonian, from which photonic bandstructures arise in a manner analogous to electronic bandstructures in a solid.  The existence of photonic bands has inspired a variety of applications in integrated optics; for instance, photonic crystals made of magneto-optical (MO) materials, which break time-reversal symmetry (T), can be used to construct nonreciprocal optical circuits \cite{Wang,Lyubchanskii}. Raghu and Haldane \cite{Haldane} have recently extended this analogy in a remarkable new direction, by predicting the existence of one-way electromagnetic modes similar to the chiral edge states found in the integer quantum Hall (QH) effect \cite{QHE}.  These modes are confined at the edge of certain 2\textsc{d} MO photonic crystals and possess group velocities pointing in only one direction, determined by the direction of an applied DC magnetic field (the time-reversed versions are not eigenstates of the system).  Due to the absence of back-propagating modes, backscattering is completely suppressed.  This is potentially important for slow-light structures, which are susceptible to backscattering \cite{Michelle2}, among other applications.  However, the requirement of locating a ``Dirac point'' in $k$-space, as suggested by Raghu and Haldane, led these authors to focus on TE modes in triangular lattices and thus to gyroelectric materials.  In realistic gyroelectric materials, the strength of T-breaking, characterized by the ratio of (imaginary) off-diagonal to (real) on-diagonal elements of the permittivity tensor $\bm{\epsilon}$ (the Voigt parameter) is at most $\sim 10^{-3}$.  As a result, the bandgap is not robust against disorder, and the edge modes scatter easily into bulk modes of the crystal, leading to significant radiative loss.  To our knowledge, these one-way edge modes have never been observed experimentally, and to do so one would desire a bandgap that is orders of magnitude broader.

\begin{figure}
\centering
\includegraphics[width=0.48\textwidth]{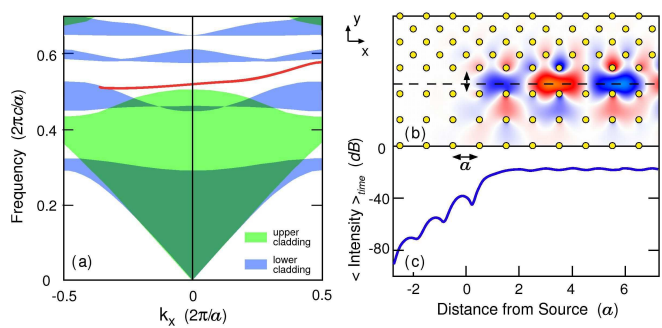}
\caption{(Color) Magneto-optical one-way waveguide. (a) Projected band diagram.  The dispersion curve (red) for the edge modes spans the band gaps for the cladding crystals. (b) Steady-state field pattern for $E_z$, the out-of-plane electric field component.  Blue and red represent positive and negative field values.  A source, indicated by double arrows, is located at the interface between a magnetized ($+z$) YIG crystal (lower half-plane) and an alumina crystal (upper half-plane), and operates at a mid-gap frequency ($0.552 \cdot 2\pi c/a$).  The excited edge mode propagates to the right and undergoes evanescent decay to the left.  (c) The time-averaged electric field intensity along the mid-line of the waveguide (dashed line in b).}
\label{Waveguide figure}
\end{figure}

In this Letter, we show that one-way modes can be generalized to photonic crystals with gyrotropic constituents without the restriction of having Dirac points in the bandstructure or the use of gyroelectric materials.  To do this, we derive an analytical mapping between the electromagnetic modes in the MO photonic crystal and the wavefunctions of a nonrelativistic electron in a QH system.  This links photonic one-way modes to QH edge states via the Hatsugai condition, which relates the number of edge modes in a band gap to the sum of Chern numbers for all bands below it \cite{Hatsugai}.  Using gyromagnetic materials, we design an experimentally-feasible one-way waveguide using a 2\textsc{d} MO square-lattice Yttrium-Iron-Garnet (YIG) photonic crystal (which lacks Dirac points) operating at microwave frequencies.  Such a one-way waveguide consists of an interface between an MO photonic crystal and a gapped material, such as a regular 2\textsc{d} photonic crystal, as shown in Fig.~\ref{Waveguide figure}.  With realistic material parameters, the one-way modes are laterally confined to a few lattice constants, and occupy a broad ($\sim 10\%$) band-gap with negligible material loss.  The MO photonic crystal exhibits strong (order unity) T-breaking due to the use of gyromagnetic ferrimagnets.  Using both frequency and time-domain numerical simulations, we demonstrate that these edge modes are immune to backscattering.  The 2\textsc{d} waveguide structure can be mapped into an equivalent 3\textsc{d} structure that supports identical field distributions and transport properties \cite{Michelle}.

For definiteness, consider an MO crystal consisting of a square lattice of YIG rods ($\epsilon=15\epsilon_0$) of radius $0.11a$ in air, where $a$ is the lattice constant. An external DC magnetic field applied in the out-of-plane ($z$) direction induces strong gyromagnetic anisotropy, with the permeability tensor taking the form
\begin{equation}
  \bm{\mu} = \begin{bmatrix}
    \mu & i\kappa & 0 \\
    -i\kappa & \mu & 0 \\
    0 & 0 & \mu_0 \\
  \end{bmatrix}.
  \label{permeability}
\end{equation}
With a 1600 Gauss applied field, the tensor elements in YIG at 4.28 GHz are $\kappa=12.4\mu_0$ and $\mu=14\mu_0$ \cite{Pozar}.  For now, we neglect the effect of material dispersion and loss, assuming a frequency-independent permeability tensor with real-valued $\kappa$ and $\mu$.  We will later show that our results are not substantially affected by these effects.

\begin{figure}
\centering
\includegraphics[width=0.38\textwidth]{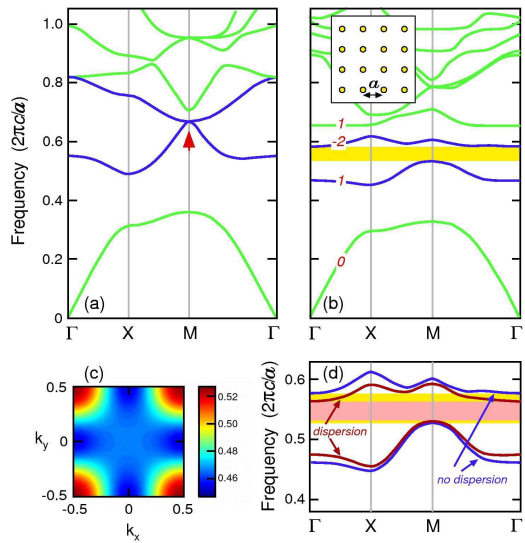}
\caption{(Color) Construction of a MO photonic crystal supporting one-way edge modes.  The crystal consists of a square lattice of YIG rods (inset in b, with $\epsilon=15\epsilon_0$ and $r = 0.11a$) in air. (a) Band diagram with zero DC magnetic field ($\mu=\mu_0$, $\kappa=0$).  The relevant quadratic degeneracy point is indicated. (b) Band diagram with a 1600 Gauss $+z$ DC magnetic field ($\mu=14\mu_0$, $\kappa=12.4\mu_0$).  The degeneracies are lifted, resulting in the given non-zero Chern numbers (red numbers).  (c) Contour plot for the second band.  Although the MO crystal has only $C_4$ symmetry, the band structure retains an accidental $C_{4v}$ symmetry, with an irreducible Brillouin zone identical to that of a non-MO crystal. The Chern numbers are calculated by integrating Eq.~\ref{Berry} along the boundary of the first Brillouin zone. (d) The corrected band gap with material dispersion.}
\label{band figure}
\end{figure}

Due to the presence of magnetic anisotropy, we must adapt the conventional band theory of photonic crystals \cite{JJ} to this system: we eliminate the \textit{magnetic} field from Maxwell's equations to obtain the master equation
\begin{equation}
  \nabla \times \left( \bm{\mu}^{-1}(\mathbf{r}) \nabla \times
  \mathbf{E} \right) = \epsilon(\mathbf{r}) \omega^2 \mathbf{E}.
  \label{Maxwell}
\end{equation}
Here, the inverse permeability tensor $\bm{\mu}^{-1}$ and the scalar permittivity $\epsilon$ are both functions of position, and $\omega$ is the mode frequency.  This equation can be cast in Hamiltonian form if we define the Hermitian inner product as
\begin{equation}
  \braket{\mathbf{E}_1|\mathbf{E}_2} = \int d^2\!r \; \epsilon(\mathbf{r})
  \mathbf{E}^*_1 \cdot \mathbf{E}_2.
  \label{inner product}
\end{equation}

The existence of one-way edge modes relies critically on T-breaking in MO crystals and its effects on the topological properties of the Bloch bands.  This effect is characterized by the Chern number, a quantity that has been extensively studied in QH systems \cite{TKNN,Simon}.  The Chern number of the $n$-th photonic band is
\begin{eqnarray}
  C_n &=& \frac{1}{2\pi i}\, \int_{BZ} d^2\!k \; \left(\frac{\partial
    \mathcal{A}_y^{nn}}{\partial k_x} - \frac{\partial
    \mathcal{A}_x^{nn}}{\partial k_y}\right),   \label{Chern} \\
  \vec{\mathcal{A}}^{nn'}(k) &\equiv& \braket{E_{nk}|\nabla_k|E_{n'k}},
  \label{Berry}
\end{eqnarray}
where the $k$-space integral in (\ref{Chern}) is performed over the first Brillouin zone, and the Bloch-function inner product in (\ref{Berry}) is defined similarly to (\ref{inner product}) with the integral performed over the unit cell.  The key properties of the Chern number are that (i) it is always an integer, (ii) the sum of the Chern numbers over all bands is zero, and (iii) the Chern number of every band is zero if the Hamiltonian is T-symmetric \cite{Simon}. Property (i) implies that as one adiabatically tunes the Hamiltonian (e.g., varying the permeability tensor by tuning the DC magnetic field bias), the Chern number of a given band changes, if and when it does so, abruptly.  In fact, this abrupt change occurs at the critical point where the band becomes degenerate with a neighboring band, at discrete ``degeneracy points'' in $k$-space.  When the system is tuned past this critical point, the degeneracies are lifted; a bandgap opens and the band's Chern number changes by an integer $\Delta C$.  (In other words, the relevant bands acquire nonzero Chern numbers as soon as a nonzero bias is applied.  However, a large bias is necessary to support a wide bandgap and a tightly-confined edge mode.)  Due to property (ii), a Chern-number change in one band is accompanied by an equal and opposite change, $-\Delta C$, in the Chern number of the neighboring band.  Hatsugai has shown that in a lattice QH system, this process creates a total of $\Delta C$ different edge states at the system's boundary
\cite{Hatsugai}.

This result, together with property (iii), implies a simple and general strategy for constructing photonic crystals supporting one-way modes.  We begin by searching for a T-symmetric band structure with a pair of photonic bands degenerate at discrete $k$-points; the requirement for Dirac points where the bands touch in a linear fashion, which was discussed by Raghu and Haldane \cite{Haldane}, is not strictly necessary.  For example, Fig.~\ref{band figure} shows the TM band diagram for a square lattice of YIG rods in air, containing a quadratic M-point degeneracy between the second and third bands.  This degeneracy is lifted upon applying an external DC magnetic field, which introduces magnetic anisotropy in the high-index rods and breaks T (as well as inversion and mirror symmetries.)  The separated bands acquire non-zero Chern numbers, causing one-way modes to appear at the edges of the sample.

The Chern number argument for the existence of photonic edge modes depends on a crucial assumption: that Hatsugai's relation between edge states and Chern numbers \cite{Hatsugai}, which was derived using a lattice QH model, applies to the photonic system.  Although this has not been formally proven, there is an interesting relationship that allows us to map the photon states in gyromagnetic crystals to electron wavefunctions in a family of QH systems.  To see this, let us return to (\ref{Maxwell}) and write $\bm{\mu}^{-1}(r)$ in terms of its component scalar functions:
\begin{equation}
  \bm{\mu}^{-1} = \begin{bmatrix}
    \tilde{\mu}^{-1} & i\eta & 0 \\
    -i\eta & \tilde{\mu}^{-1} & 0 \\
    0 & 0 & \mu_0^{-1} \\
  \end{bmatrix}.
  \label{inverse permeability}
\end{equation}
In terms of the component functions in (\ref{permeability}), $\tilde{\mu}^{-1} = \mu / (\mu^2 - \kappa^2)$ and $\eta = - \kappa / (\mu^2 - \kappa^2)$.  The equations for $E_x$ and $E_y$ decouple from $E_z$; for TM states $(E_x = E_y = 0)$,
\begin{equation} \left[ - \nabla^2 +
    \left(\nabla\ln\tilde{\mu} - i \tilde{\mu} \hat{z} \times \nabla \eta\right)
    \cdot \nabla - \tilde{\mu} \epsilon\omega^2 \right] E_z = 0.
\end{equation}
Expressing this in terms of $\psi \equiv E_z\sqrt{\tilde{\mu}}$, we obtain
\begin{equation}
  \left[ - \left|\nabla + i \tilde{A}(r) \right|^2 + \tilde{V}(r)
    \right] \psi = 0,
\end{equation}
where
\begin{eqnarray}
  \tilde{A} &=& \frac{\tilde{\mu}}{2}\; \hat{z}\times \nabla\eta, \\
  \tilde{V} &=& \frac{1}{4} \left(|\nabla\ln\tilde{\mu}|^2 +
  |\tilde{\mu}\nabla\eta|^2\right) - \frac{1}{2} \nabla^2\ln\tilde{\mu} -
  \tilde{\mu}\epsilon\omega^2.
  \label{V}
\end{eqnarray}
This is the equation for zero-energy wavefunctions of a non-relativistic particle in periodic vector and scalar potentials $\tilde{A}(r)$ and $\tilde{V}(r)$.  Increasing $\omega$ corresponds to increasing the depth of the scalar potential well in the third term of (\ref{V}), shifting the spectrum downwards relative to the zero of energy.  With suitable boundary conditions, this mapping holds for both edge modes and bulk modes.  Thus, for each value of $\omega$, the existence of unpaired edge modes, as well as their spatial characteristics, can be mapped to a similar problem in a QH system.

To complete the single-mode one-way waveguide, we interface the MO crystal with a upper cladding that supports no bulk modes at the frequency of the second MO crystal bandgap. This can be achieved using a metal slab, or using a regular photonic crystal with an aligned bandgap.  The topological nature of the edge modes \cite{Haldane} ensures that the interface structure is unimportant, as long as it is sufficiently narrow that higher order modes are avoided.  Here, we use a square lattice of high-index alumina rods ($r=0.106a, \epsilon=10\epsilon_0$) in air, tilted 45 degrees to match the bandgap frequency (Fig.~\ref{Waveguide figure}b).  Note that the regular crystal has zero Chern numbers for all its photonic bands, so there is only one forward-propagating mode at the interface.  With a 1600 Gauss bias, we obtain a wide MO bandgap of 10\% relative size and very low dispersion, with the edge mode confined to a few lattice constants. Using a finite-element frequency-domain scheme and a steady-state excitation, we observe an evanescent mode decaying exponentially along the backward direction of the waveguide (Fig.~\ref{Waveguide   figure}c).

\begin{figure}
\centering
\includegraphics[width=0.276\textwidth]{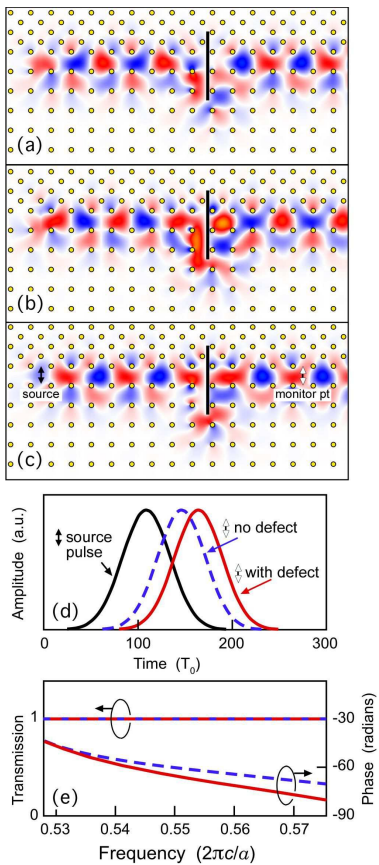}
\caption{(Color) Back-scattering suppression in the one-way waveguide.  Inserting a slab of perfect electrical conductor with thickness $0.2a$ (black rectangle) causes the propagating modes to circumvent the defect, maintaining complete transmission.  $E_z$ is plotted for (a) $t=0$, (b) $t=0.25 T_0$, and (c) $t=0.5T_0$, where $T_0$ is the optical period.  (d) Time-domain simulation results for a temporal Gaussian pulse with spectrum contained in the bandgap.  The electric field amplitude is plotted at the source point (black), at the same transverse position 13 lattice constants downstream along the waveguide in the absence of a defect (red), and at the same point with an intervening defect (blue).  The pulse is completely transmitted regardless of the existence of the defect.  (e) Transmission and phase shift plots for the time-domain simulation.}
\label{Scattering figure}
\end{figure}

One of the unusual properties of the one-way edge modes is the complete suppression of backscattering, which is an optical analog of the dissipation-free transport of edge electrons in QH systems. Numerical simulations show that these edge modes are immune to scattering from extremely large defects.  For instance, Fig.~\ref{Scattering figure} shows the results of simulations with a slab of perfect electrical conductor (PEC) of width 3$a$ and thickness 0.2$a$ inserted into the waveguide.  In a conventional waveguide, such a drastic defect would almost completely block the guided mode.  In the one-way waveguide, a steady-state source operating at the mid-gap frequency $0.555 \cdot (2\pi c/a)$ excites a one-way mode that circumvents the PEC defect, with 100\% power transmission throughout the MO bandgap.  This happens because the defect creates a new interface waveguide between the PEC and the MO crystal.  Thus, it only alters the phase response, which is partly due to the delay incurred by traversing the lengthened interface.  We corroborated this result with a time-domain calculation in which a temporal Gaussian pulse with a spectral bandwidth of 50\% of the bandgap and carrier frequency $0.555 \cdot (2\pi c/a)$ is launched into the waveguide. Regardless of the presence of the defect, the pulse passes through the waveguide with no perceivable change in amplitude or pulse width; since the one-way mode has approximately linear dispersion relation at mid-gap frequencies (Fig.~\ref{Waveguide figure}), sharp corners do not contribute significantly to chromatic dispersion.  The increased transit time is in agreement with the change in group delay.  Although embedded sources were used in these simulations, we have verified in a separate set of simulations that it is also easy to couple to the one-way mode using waves incident on the system boundary.

While our analysis has thus far been limited to 2\textsc{d} structures at microwave frequencies, similar one-way waveguides can be realized in practical 3\textsc{d} structures and at higher frequencies. In the 2\textsc{d} analysis, the dielectric structure and electromagnetic field extend uniformly along the $z$ direction.  In a 3\textsc{d} structure, TM modes can be truncated by introducing a PEC (for which a metal slab is a good approximation at microwave frequencies) in the $x$-$y$ plane, without affecting the field distribution.  An array of finite high YIG rods and alumina rods between two metal slabs supports TEM modes with field distributions identical to the TM modes in our 2\textsc{d} system \cite{Michelle,Pozar}.  Finally, although gyromagnetic effects are generally limited to the GHz range by the availability of high-field magnetic bias, artificial magnetic resonance, such as those exploited in metamaterials, might be incorporated to synthesize the gyrotropic responses at THz or even infrared frequencies \cite{Pendry,Zhang,Wegener}.

The above results are not substantially altered by the material losses and frequency-dependent permeability in a real microwave ferrite. With a gyromagnetic linewidth of 0.3 Oe and a dieletric loss tangent of 0.0002, typical in commerically available monocrystalline YIG \cite{Wohlfarth}, the complex propagation constant is $(-0.359+0.0001i)(2\pi/a)$ in the case of Fig.~3.  The imaginary part corresponds to a decay length of 300 lattice constants, far exceeding practical structural dimensions.  We have also neglected the frequency dependency of the gyromagnetic permeability.  When correcting this approximation with a complete gyrotropic lineshape \cite{Pozar}, we find the MO bandgap reduces to around 6\%. Although such reduction narrows the operational bandwidth, we find the dispersion has no impact on the back-scattering suppresion or the confinement of the edge mode. We envision many applications for these one-way waveguides, including defect-tolerant slow-light systems \cite{Michelle2}, microwave isolators, high-Q channel add/drop filters, and all-pass filters.  It might also be possible to further exploit the analogy between this system and the QH effect, by for instance reproducing the fractal ``Hofstadter butterfly'' spectrum \cite{Hofstadter} in the bulk modes of the MO photonic crystal.

We thank S.~Fan, J.~T.~Shen, F.~D.~M.~Haldane, and X.~G.~Wen for useful discussions, and Y.~Fink for his encouragement and support. This research was supported in part by the Army Research Office through the Institute for Soldier Nanotechnologies under Contract No.~DAAD-19-02-D0002, and in part by the Materials Research Science and Engineering Center program of the National Science Foundation under Grant No.~DMR 02-13282.


\begin{thebibliography}{99}

\bibitem{Sajeev}
  S.~John, \textit{Phys.~Rev.~Lett.}~{\bf 58}, 2486 (1987).

\bibitem{Yablonovitch}
  E.~Yablonovitch, \textit{Phys.~Rev.~Lett.}~{\bf 58}, 2059 (1987).

\bibitem{JJ}
  J.~D.~Joannopoulos, R.~D.~Meade, and J.~N.~Winn,
  \textit{Photonic Crystals: Molding the Flow of Light}
  (Princeton University Press, Princeton, 1995).

\bibitem{Wang}
  Z.~Wang, and S.~F.~Fan, \textit{Opt.~Lett.}~{\bf 30}, 1989 (2005).

\bibitem{Lyubchanskii}
  I.~L.~Lyubchanskii \textit{et.~al.}, \textit{J.~Phys.~D}~{\bf 36}, R277 (2003).

\bibitem{Haldane}
  Raghu, S.~and Haldane, F.~D.~M.
  cond-mat/0503588 and cond-mat/0602501.

\bibitem{QHE}
  R.~E.~Prange and S.~M.~Girvin, ed. \textit{The Quantum Hall Effect}.
  (Springer-Verlag, New York, 1987).

\bibitem{Michelle2}
  M.~L.~Povinelli \textit{et.~al.}, \textit{Appl.~Phys.~Lett.}~{\bf 84}, 3639 (2004).

\bibitem{Hatsugai}
  Y.~Hatsugai, \textit{Phys.~Rev.~Lett.}~{\bf 71}, 3697 (1993).

\bibitem{Michelle}
  M.~L.~Povinelli \textit{et.~al.}, \textit{Phys. Rev.}~{\bf B64}, 075313 (2001).

\bibitem{Pozar}
  D.~M.~Pozar, \textit{Microwave Engineering} (John Wiley, New York, 1998).

\bibitem{TKNN}
  D.~Thouless \textit{et.~al.}, \textit{Phys.~Rev.~Lett.}~{\bf 49}, 405 (1982).

\bibitem{Simon}
  B.~Simon, \textit{Phys.~Rev.~Lett.}~{\bf 51}, 2167 (1983).

\bibitem{Pendry}
  J.~B.~Pendry \textit{et.~al.}, \textit{IEEE Trans. Microwave Theory Tech.}~{\bf 47}, 2075 (1999).

\bibitem{Zhang}
  T.~J.~Yen \textit{et.~al.}, \textit{Science}~{\bf 303}, 1494-1496 (2004).

\bibitem{Wegener}
  S.~Linden \textit{et.~al.}, \textit{IEEE J. Sel. Top. Quantum Electron.}~{\bf 12}, 1097 (2006).

\bibitem{Wohlfarth}
  E.~P.~Wohlfarth, in \textit{Ferromagnetic Materials }, Vol.~{\bf 2}, 293 (North-Holland, Amsterdam, 1986).

\bibitem{Hofstadter}
  D.~Hofstadter, \textit{Phys.~Rev.} {\bf B14}, 2239 (1976).

\end{thebibliography}
\end{document}